\documentclass[journal]{IEEEtran}

\usepackage[T1]{fontenc}
\usepackage[utf8]{inputenc}
\usepackage[greek,english]{babel}
\usepackage{cite}
\usepackage{scalerel}
\usepackage{enumerate}
\usepackage{graphicx}
\usepackage[export]{adjustbox}
\usepackage{multirow}
\usepackage{verbatim}
\usepackage{tikz}
\usepackage{float}
\usepackage{setspace}
\usepackage{time}
\usepackage{xspace}
\usepackage{xcolor}
\usepackage{amsmath,amssymb,algorithm,hyperref}
\usepackage{subcaption}
\usepackage{mwe}

\newtheorem{mydef}{Definition}

\def\BibTeX{{\rm B\kern-.05em{\sc i\kern-.025em b}\kern-.08em
    T\kern-.1667em\lower.7ex\hbox{E}\kern-.125emX}}

\begin{document}

\title{Low Complexity Optimization of the Asymptotic Spectral Efficiency in Massive MIMO NOMA 
}

\author{Lucinda Hadley and Ioannis Chatzigeorgiou%
\thanks{This work has been funded by Lancaster University's EPSRC Doctoral Training Partnership.

L. Hadley and I. Chatzigeorgiou are with InfoLab21, School of Computing and Communications, Lancaster University, LA1~4WA, UK (e-mail: \{lucinda.hadley, i.chatzigeorgiou\}@lancaster.ac.uk).}%
}
\maketitle
\begin{abstract}
Massive multiple-input multiple-output (MIMO) technology facilitates huge increases in the capacity of wireless channels, while non-orthogonal multiple access (NOMA) addresses the problem of limited resources in traditional orthogonal multiple access (OMA) techniques, promising enhanced spectral efficiency. This work uses asymptotic capacity computation results to reduce the complexity of a power allocation algorithm for small-scale MIMO-NOMA, so that it may be applied for systems with massive MIMO arrays. The proposed method maximizes the sum-capacity of the considered system, subject to power and performance constraints, and demonstrates greater accuracy than alternative approaches despite remaining low-complexity for arbitrarily large antenna arrays.
\end{abstract}


\section{Introduction}
The demand for fast data links has increased rapidly over the last two decades as a result of an increasing number of users and devices. Moreover, there is a need for adaptable and scalable technologies to meet the diverse requirements of the internet of things (IoT). Fifth generation (5G) and sixth generation (6G) networks must be able to support increased multi-terabyte per second data traffic, while maintaining a high quality of service in terms of  security, reliability and delay \cite{zhang20196g}.

A key facilitator of the increased spectral efficiency (SE) seen between third and fourth generation mobile networks was the use of  multiple-input multiple-output (MIMO) technology. MIMO enables dramatic increases in SE by exploiting spatial diversity \cite{hampton2013introductiontoMIMO} and can be extended by using even more antennas in `massive MIMO' (MM). In 2018, a line of products with MM capability was approved by the Federal Communications Commission. These included 64-antenna arrays, such as the Ericsson AIR 6468. Similar products, including the Huawei AAU and Nokia Airscale, have also been launched with Huawei quoted as saying at the 2019 Mobile World Congress that ``95\% of their current commercial shipments has either 32 or 64 antennas'' \cite{bjornson2019massive}. It is speculated that antenna arrays with dimensions of order $10^3$ or even $10^4$ could be used in future designs in so called `supermassive MIMO'. MM is therefore of critical importance in industry and large-scale arrays are a topic of great interest in current research \cite{zhang20196g}.

Rate optimization of a wireless network requires knowledge of the theoretical SE of its channels. In 1999 Telatar's groundbreaking work introduced the use of asymptotic properties of random matrices, in particular the limiting distributions of their eigenvalues, in computing the asymptotic SE of MIMO channels \cite{telatar1999capacity}. In 2004, \cite{tulino2004random} and \cite{muller2004random} demonstrated some ways of generalizing the result, but the work was premature with respect to small-scale MIMO, whose capacity is more easily computed using the celebrated `log-det' result \cite{biglieri2007mimo}. With the recent introduction of MM, however, the analysis of very large random matrices is required, and the use of asymptotic results has resurfaced. The last several years have seen methods, such as free probability theory, used to compute the asymptotic eigenvalue distributions (AEDs) of a wider class of MIMO channel matrices \cite{pan2013asymptotic, diaz2017capacity, hadley2019}.

Another method for enhancing SE is to share spectrum more effectively. Non-orthogonal multiple access (NOMA) is an emerging technology that shows promise in this area. Traditional NOMA uses the power domain to discriminate between signals (although a code-domain implementation of NOMA has also been proposed) \cite{vaezi2019multiple}. Unlike orthogonal multiple access (OMA) methods, such as time and frequency division multiple access (TDMA and FDMA), which split the respective resources (spectrum and time) into `orthogonal' frequency bands and time slots, NOMA serves multiple users in a single resource block (band or slot), thus enabling massive connectivity. This, along with the mitigating effect of using successive interference cancellation (SIC) to remove unwanted signals and improve the signal-to-interference-plus-noise ratio (SINR), results in increased SE \cite{ding2014performance}. NOMA is considered fairer than alternative multiple access schemes as it prioritizes the experience of cell-edge users with weaker channel connections. Moreover, it reduces average latency compared to OMA since users do not have to wait for specific slots \cite{islam2017noma}. 

Due to early results demonstrating its potential, NOMA already features in the 3GPP-LTE-A standard and was proposed for inclusion in the 5G New Radio (NR) \cite{ding2017application}. Ultimately, NOMA was not included in 5G NR as a work-item, but was earmarked for use beyond 5G because the capacity benefits were considered to be outweighed by the implementation complexity \cite{makki2020survey, 3GPPspecNOMA}. Therefore, it is necessary to increase the capacity benefits in relation to the complexity in order to make NOMA a viable option, and the use of massive antenna arrays is an obvious strategy. For the multi-user case in which the base station is equipped with multi-antenna arrays, while the user devices have a single antenna, \cite{islam2018resource} compares some user-pairing algorithms and investigates a new method for maximizing throughput, while in \cite{zeng2017capacity} the authors demonstrate the superior capacity of MIMO-NOMA over MIMO-OMA for communication between a multi-antenna receiver and clusters of multi-antenna destinations. This is extended to massive-MIMO NOMA (MM-NOMA) in \cite{zhang2017performance}, which shows that a non-regenerative relay system where the base station is equipped with up to $500$ antennas, outperforms a traditional MIMO-NOMA arrangement.

In this work we consider a low-complexity power allocation algorithm for two-user power-domain NOMA in which MM arrays are employed at all nodes and signals can be separated using superposition coding (SC) at the transmitter and SIC at the receiver. We assume that the transmitter has access to statistical channel state information (CSIT) only and we aim to maximize the ergodic capacity subject to power and rate constraints. This non-convex optimization problem was addressed for the case of small-scale MIMO by implementing a suboptimal algorithm and comparing it to the optimal bisection method in \cite{sun2015ergodic}. We extend the work to consider arbitrarily large MM arrays and demonstrate that it is possible to reduce the complexity of the bisection method further, by combining it with Telatar's method of asymptotic capacity computation, without loss of optimality. As far as the authors are aware, this approach has not previously been considered for this scenario.

\textit{Notations:} $(\cdot)^\dag$ denotes the conjugate transpose, $\mathrm{Tr}(\cdot)$ represents the matrix trace, $\mathbb{I}_N$ denotes the $N \times N$ identity matrix and $\mathbb{E}(\cdot)$ is the expectation.

%

\section{System Model}

\begin{figure}[t]
\centering
\includegraphics[width=0.67\columnwidth]{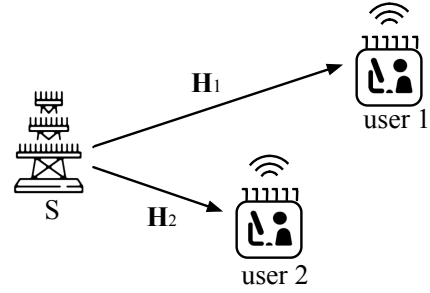}
\caption{Broadcast MM-NOMA system model using SIC.}
\label{figsystemmodel}
\end{figure}

Consider the open-loop MIMO system given in Fig. \ref{figsystemmodel}, where a source $\mathrm{S}$ transmits data to two users simultaneously using $N_\mathrm{S}$ antennas and user $i$ receives using $N_i$ antennas, where $i\in\{1,2\}$. The signal vectors $\mathbf{x}_1$ and $\mathbf{x}_2$ are transmitted to user~1 and user~2 and the diagonal power allocation matrices are $\mathbf{Q}_1,\mathbf{Q}_2 \in \mathbb{C}^{N_\mathrm{S}\times N_\mathrm{S}}$ at each user respectively, where $\mathrm{Tr}(\mathbf{Q}_i)$ is the total power allocated to user $i$. Both signals occupy the same frequency and time slot but their transmit power varies, as is the usual convention for NOMA transmission. User~1 and user~2 are taken to be the `weak user' and `strong user', respectively. This could occur, for example, when $\mathrm{S}$ is a base station, user~1 is at the cell-edge and user~2 is near the center of the cell. It was determined in \cite[Lemma~2]{sun2015ergodic} that uniform power allocation across each user's antennas results in optimal performance. Therefore, hereafter we will consider the case where the diagonal entries of $\mathbf{Q}_i$ are all equal and replace each $\mathbf{Q}_i$ with the constant scalar $p_i=\frac{\mathrm{Tr}(\mathbf{Q}_i)}{N_\mathrm{S}}$, which represents the power allocated to the desired signal of user $i$ per antenna at the source. 

User~1 and user~2 receive signals $\mathbf{y}_1$ and $\mathbf{y}_2$ respectively, which can be expressed as:
\begin{align}
\mathbf{y}_1 &=\sqrt{p_1}\,\mathbf{H}_1\mathbf{x}_1 + \sqrt{p_2}\,\mathbf{H}_1\mathbf{x}_2 + \mathbf{n}_1,\\
\mathbf{y}_2 &=\sqrt{p_1}\,\mathbf{H}_2\mathbf{x}_1 + \sqrt{p_2}\,\mathbf{H}_2\mathbf{x}_2 + \mathbf{n}_2,
\end{align}
where $\mathbf{x}_i$ is the $N_\mathrm{S} \times 1$ vector of the transmitted signal carrying the message for user $i$, and $\mathbf{y}_i$ is the $N_i \times 1$ vector of the signal received by user $i$. Matrices $\mathbf{H}_i\in \mathbb{C}^{N_i \times N_\mathrm{S}}$ have random complex entries distributed as $\mathcal{CN}(0,\sigma^2_{\mathbf{H}_i})$, which model flat Rayleigh fading. Each entry of $\mathbf{H}_i$, denoted by $h_{jk}^i$, represents the channel gain between the $k$th transmit antenna of $\mathrm{S}$ and the $j$th receive antenna of user~$i$. We assume that $\sigma^2_{\mathbf{H}_1}<\sigma^2_{\mathbf{H}_2}$ because user~1 is the weak user. Finally, the $N_i\times 1$ vector $\mathbf{n}_i$ models the normalized additive white Gaussian noise across the corresponding channel. 

Since we are using NOMA, the source simultaneously communicates with the users using the same resource block, and their signals are multiplexed by allocating a different transmission power, $p_i$, for each user's signal, at each antenna. Because the weaker user is allocated more power, it is able to decode the message by treating the interference from the other user's signal as noise. Define $C_1$ and $C_2$ as the SEs of user~1 and user~2 respectively. We will set a minimum rate constraint of $C_1>R_0$ for the weak user and assume that the SINR of the weak user's signal is always smaller at the weak user than it is at the strong user so that
\begin{equation}
\label{eqnlimit}
C_1\le\log_2\left|\mathbb{I}_{N_2}+\left(\mathbb{I}_{N_2}+p_2\mathbf{H}_2\mathbf{H}_2^\dag\right)^{-1}p_1\mathbf{H}_2\mathbf{H}_2^\dag\right|,
\end{equation}
which guarantees successful SIC detection at the strong user. This means that the strong user can decode the weak user's message and subtract it from the overall signal in order to decode its own message \cite{ding2015application2}.

The weak user decodes its own signal, $\mathbf{x}_1$, while interpreting the interference caused by $\mathbf{x}_2$ as noise. The achievable ergodic SEs are therefore given by:
\begin{align}
C_1 =& \mathbb{E}_{\mathbf{H}_1}\left(\log_2 \left|\mathbb{I}_{N_1}+\left(\mathbb{I}_{N_1}+p_2\mathbf{H}_1\mathbf{H}_1^\dag\right)^{-1}p_1\mathbf{H}_1\mathbf{H}_1^\dag\right|\right) \nonumber \\
=& \mathbb{E}_{\mathbf{H}_1}\left(\log_2 \left|\mathbb{I}_{N_1}+\left(p_1+p_2\right)\mathbf{H}_1\mathbf{H}_1^\dag \right| \right) \nonumber\\
&\quad\quad\quad\quad\quad -\mathbb{E}_{\mathbf{H}_1}\left(\log_2 \left|\mathbb{I}_{N_1}+p_2\mathbf{H}_1\mathbf{H}_1^\dag \right| \right), \label{eqnc1} \\
C_2 =& \mathbb{E}_{\mathbf{H}_2}\left(\log_2 \left|\mathbb{I}_{N_2}+p_2\mathbf{H}_2\mathbf{H}_2^\dag \right| \right). \label{eqnc2}
\end{align}

\section{Optimization Problem}
\label{secop}

The optimization problem of maximizing the combined SE of the two users, subject to power and minimum rate constraints, can be formulated as:
\begin{equation}
\label{pa_opt}
\begin{aligned}
\max_{p_1, p_2 \ge 0}\quad 
	&\begin{array}{l}
	C_1(p_1,p_2)+C_2(p_2),
	\end{array} \\
\textrm{s.t.}\quad
   &\begin{array}{r@{}l}
    {}&C_1(p_1,p_2) \ge R_0 \\
    {}&(p_1+p_2)N_\mathrm{S} \le p_{\textrm{max}},
    \end{array}
\end{aligned}
\end{equation}
where $p_{\textrm{max}}$ denotes the total available power at the source, $R_0$ is the minimum SE required for reasonable performance at the weak user and $C_1(p_1,p_2)$ and $C_2(p_2)$ refer to the SEs defined in \eqref{eqnc1} and \eqref{eqnc2} respectively, written in terms of the optimization variables $p_1$ and $p_2$.

\begin{table}
\begin{center}
\caption{Optimal bisection algorithm$^\ddagger$}
\label{tab1}
\begin{tabular}{|l|}
\hline
\texttt{Initialize p}$_{\texttt{2,min}}$ \texttt{=} \texttt{0, p}$_{\texttt{2,max}}$\texttt{ = p}$_\texttt{max}$\\
\hline
\texttt{\textbf{while} p}$_{\texttt{2,max}}$\texttt{ - p}$_{\texttt{2,min}}$ $\texttt{> } \textgreek{\epsilon}$ \texttt{\textbf{do}}\\
\texttt{Set p}$_{\texttt{2}}^{\texttt{*}}$\texttt{ = (p}$_{\texttt{2,min}}$\texttt{ + p}$_\texttt{2,max}$\texttt{)/2,}\\
\qquad\:\! \texttt{p}$_\texttt{1}^\texttt{*}$\texttt{ = p}$_\texttt{max}$\texttt{ - p}$_\texttt{2}^\texttt{*}$\texttt{.}\\
\texttt{Calculate C}$_\texttt{1}$\texttt{(p}$_\texttt{1}^\texttt{*}$\texttt{,p}$_\texttt{2}^\texttt{*}$\texttt{).}\\
\texttt{If C}$_\texttt{1}$\texttt{(p}$_\texttt{1}^\texttt{*}$\texttt{,p}$_\texttt{2}^\texttt{*}$\texttt{)< R}$_\texttt{0}$\texttt{, set p}$_\texttt{2,max}$ \texttt{= p}$_\texttt{2}^\texttt{*}$;\\
\qquad \texttt{Else, set p}$_\texttt{2,min}$ \texttt{= p}$_\texttt{2}^\texttt{*}$\texttt{.}\\
\texttt{\textbf{end while}}\\
\hline
\texttt{Output: p}$_\texttt{1}$ \texttt{=} \texttt{p}$_\texttt{1}^\texttt{*}$\texttt{, p}$_\texttt{2}$ \texttt{=} \texttt{p}$_\texttt{2}^\texttt{*}$\texttt{.}\\
\hline
\end{tabular}\\
\vspace{2 px}
\small{$^\ddagger$ \texttt{p}$_\texttt{max}$ in the algorithm is set equal to $p_{\textrm{max}}/N_\mathrm{S}$ as per \eqref{pa_opt}.}
\end{center}
\end{table}

In \cite{sun2015ergodic} the authors develop an optimal and suboptimal method of solving the problem. Since the function $C_1+C_2$ increases with $p_2$, the optimal solution is on the boundary of the feasible region. In particular, it occurs when $p_1$ is as small as possible while ensuring that $C_1>R_0$. This $p_1$ can be found using repeated bisection as shown in Table \ref{tab1}, where $\epsilon$ is reduced for greater precision. The suboptimal method relies on an approximation of $C_1$ and is successful for MIMO systems with $N_\mathrm{S}, N_i\le4$. However, the optimality of the results using this method deteriorates as the numbers of antennas at each end of the communication link increase. 

In this paper, we demonstrate how to reduce the complexity of the optimal bisection method by computing $C_1$ using the asymptotic eigenvalue distribution of the channel matrices, thus improving the accuracy of the optimization for \mbox{MM-NOMA} systems.

\section{Theory}
Let $\mathbf{G}_{\beta}\in\mathbb{C}^{N_r\times N_t}$ be a random matrix, where the limit of the ratio ${\frac{N_t}{N_r}}$ is $\beta$ as both $N_t$ and $N_r$ tend to infinity, and ${\mathbf{X}_{\beta}=\mathbf{G}_{\beta}\mathbf{G}_{\beta}^\dag\in\mathbb{C}^{N_r \times N_r}}$. When the entries of $\mathbf{G}_{\beta}$ conform to certain distribution rules and $\alpha$ is a scalar, a `log-det' expression, $\frac{1}{N_r}\log_2\left|\mathbb{I}_{N_r}+\alpha\mathbf{X}_{\beta}\right|$ can be expressed in terms of the AED, $f_{\mathbf{X}_{\beta}}(x)$, of $\mathbf{X}_{\beta}$. Using this result, the SE of a channel modeled as $\mathbf{G}_{\beta}$ can then be written in terms of the AED of $\mathbf{X}_{\beta}$ as \cite{tulino2004random}:
\begin{align}
\mathcal{C}^{Asy}_{\alpha\mathbf{X}_{\beta}}&=N_r\left(\displaystyle\lim_{\substack{N_t,N_r\rightarrow \infty \\ \frac{N_t}{N_r}\rightarrow \beta}} \displaystyle\frac{1}{N_r}\log_2 \left| \mathbb{I}_{N_r} + \alpha\mathbf{X}_{\beta}\right|\right) \nonumber \\
&=N_r\left(\displaystyle\lim_{\substack{N_\mathrm{t},N_r\rightarrow \infty \\ \frac{N_r}{N_\mathrm{t}}\rightarrow \beta}}  \displaystyle\frac{1}{N_r}\sum_{i=1}^{N_i} \log_2\left(1+ \alpha\lambda_{\mathbf{X}_{\beta}}(i)\right)\right) \nonumber \\
\label{eqncapeigs}
&= N_r \displaystyle\int_0^\infty\, \log_2\left(1+ \alpha x\right)f_{\mathbf{X}_{\beta}}(x)\,dx,
\end{align}
where $\lambda_{\mathbf{X}_{\beta}}(i)$ is the $i$th eigenvalue of $\mathbf{X}_{\beta}$. 

There are many existing works in which the main result has been to compute the AEDs of non-standard channel matrices, usually with the aim of applying \eqref{eqncapeigs} to compute their capacity. For example, Pan \textit{et al.} \cite{pan2013asymptotic} use free probability theory to compute the AED of massive MIMO channel matrices with transmit and receive correlation. Hadley \textit{et al.} \cite{hadley2019} derive the AED of the combined channels in the second hop of a multi-relay system, while Diaz and P{\'e}rez-Abreu \cite{diaz2017capacity}  find the AED for more generalized block matrices. Shlyakhtenko \cite{shlyakhtenko1996random} shows how to extend existing results to find the AED of band Gaussian matrices used to model independent but non-identically distributed Gaussian channels. 

In this paper, the channels are modeled as having entries distributed as $\mathcal{CN}(0,\sigma^2_{\mathbf{H}_i})$, which is the canonical model for single-user narrowband MIMO channels \cite{tulino2004random}, and so we make use of the following result.

\begin{mydef}
\label{defMP}
\textit{
The AED of $\mathbf{X}_\beta=\mathbf{G}_{\beta}\mathbf{G}_{\beta}^\dag$ as $N_t, N_r \rightarrow \infty$ and $\frac{N_t}{N_r}\rightarrow \beta$, where $\mathbf{G}_{\beta}\in \mathbb{C}^{N_r\times N_t}$ is a standard Gaussian random matrix with entries distributed as $\mathcal{CN}(0,1)$, is given by the Mar\c{c}enko-Pasteur distribution \cite{tulino2004random}: }
\begin{equation}
\label{marc}
f_{\mathbf{X}_{\beta}}(x)=\frac{\sqrt{\left(x-a\right)^+\left(b-x\right)^+}}{2\pi\beta x}+\left(1-\frac{1}{\beta}\right)^+\delta(x),
\end{equation}
\textit{where $a=\left(1-\sqrt{\beta}\right)^2$, $b=\left(1+\sqrt{\beta}\right)^2$, $(z)^+ = \max(0, z)$ and $\delta(x)$ is the Dirac-delta function. }
\end{mydef}

Our aim is to find $C_1$ and $C_2$, as given in \eqref{eqnc1} and \eqref{eqnc2}. Now our channel matrices $\mathbf{H}_i$ can be written as $\sigma_i\mathbf{G}_{\beta_i}$, where we have substituted $\beta=\beta_i$ into Definition \ref{defMP} so that $\mathbf{H}_i\mathbf{H}_i^\dag=\sigma_i^2\mathbf{X}_{\beta_i}$. Therefore, to find $C_1$ and $C_2$ in closed form, we can apply \eqref{eqncapeigs} to obtain:  
\begin{align}
C_1&=\log_2\left|\mathbb{I}_{N_1}+  c_1\mathbf{X}_{\beta_1} \right| - \log_2\left| \mathbb{I}_{N_2} +  c_2\mathbf{X}_{\beta_1}\right|\nonumber\\
&= \mathcal{C}^{Asy}_{c_1\mathbf{X}_{\beta_1}} - \mathcal{C}^{Asy}_{c_2\mathbf{X}_{\beta_1}} \nonumber \\
&= \displaystyle\int_0^\infty\, \log_2\left( \frac{{1+ c_1 x}}{{1+ c_2 x}}\right)^{f_{\mathbf{X}_{\beta_1}}(x)} \,dx \nonumber\\
&= \log_{2}\!\left(\!\frac{e^{\frac{\mathcal{Q}_{2,1}}{c_2}}\!\left(1+c_1-\mathcal{Q}_{1,1}\right)^{\beta_1}\!\left(1+c_1\beta_1-\mathcal{Q}_{1,1}\right)}{e^{\frac{\mathcal{Q}_{1,1}}{c_1}}\!\left(1+c_2-\mathcal{Q}_{2,1}\right)^{\beta_1}\!\left(1+c_2\beta_1-\mathcal{Q}_{2,1}\right)}\!\right)\! \label{eqnc1asy}\\
C_2&=\mathcal{C}^{Asy}_{c_3\mathbf{X}_{\beta_2}}\nonumber \\
&=\displaystyle\int_0^\infty\, \log_2\left(1+ c_3 x\right)f_{\mathbf{X}_{\beta_2}}(x)\nonumber\\
&= \log_{2}\!\left(\frac{\left(1+c_3-\mathcal{Q}_{3,2}\right)^{\beta_2}\left(1+c_3\beta_2-\mathcal{Q}_{3,2}\right)}{e^{\frac{\mathcal{Q}_{3,2}}{c_3}}}\right),\label{eqnc2asy}
\end{align}
where  $c_1=(p_1+p_2)\sigma_1^2$, $c_2=p_2\sigma_1^2$, $c_3=p_2\sigma_2^2$, $f_{\mathbf{X}}(x)$ is given by \eqref{marc} and, for notational convenience, we have set:
\[
\mathcal{Q}_{\rho,q}=\frac{1}{4}\left(\!\sqrt{c_\rho\left(1+\sqrt{\beta_q}\right)^{\!2} +1}-\sqrt{c_\rho\left(1-\sqrt{\beta_q}\right)^{\!2} +1}\right)^{\!2}\!\!.
\]

\section{Results and Discussion}
In this section we compare: (i) the bisection algorithm described in \cite{sun2015ergodic}, which relies on the traditional method of capacity computation given in \eqref{eqnc1} and \eqref{eqnc2} and finds the optimal power allocation, (ii) the suboptimal algorithm also derived in \cite{sun2015ergodic} which omits the need for repeated bisections but still relies on computing the expectation over multiple realizations of the determinant of a matrix, and (iii) the bisection method using our asymptotic capacity equations \eqref{eqnc1asy} and \eqref{eqnc2asy} in placec of the traditional method. For the sake of simplicity, we have considered the cases where $N_\mathrm{S}=N_i=N$ in our results.

\begin{figure}[tbp]
\centering
\includegraphics[width=\columnwidth]{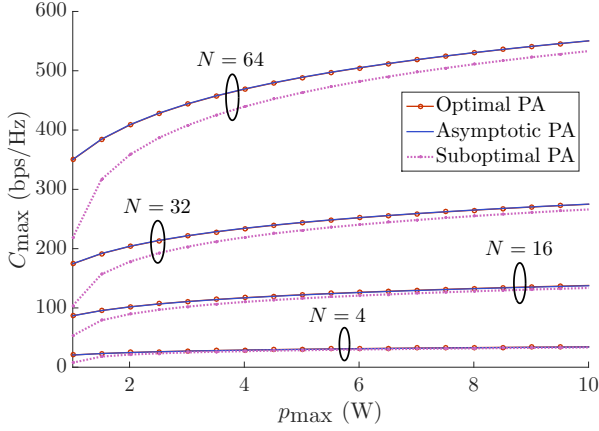}
\caption{Sum-capacity vs total transmission power}
\label{pmaxvscap}
\end{figure}

Fig.~\ref{pmaxvscap} plots the total available power $p_{\text{max}}$ against the maximized sum of the ergodic capacities of the two users obtained using \eqref{pa_opt}, which we shall denote by $C_{\textrm{max}}$. We fixed $\sigma^2_{\mathbf{H}_1}\!=\!20$~dB, $\sigma^2_{\mathbf{H}_2}\!=\!5$~dB and $R_0\!=\!2$~bps/Hz. Both the asymptotic and suboptimal methods appear to achieve very close to optimal performance for smaller MIMO arrays of $4\times4$ antennas, however, as we increase the number of antennas the suboptimal method becomes less efficient. On the other hand, the asymptotic approach is able to match the optimal result perfectly regardless of the array size. The suboptimal result is also shown to be less accurate for systems with low power availability, while the asymptotic approach is unaffected.

%

\begin{figure}[tbp]
\centering
\includegraphics[width=\columnwidth]{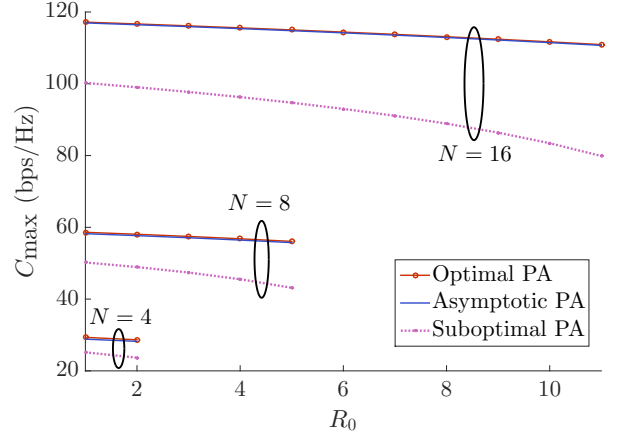}
\caption{Sum-capacity vs minimum rate of weak user}
\label{R2vscap}
\end{figure}

Fig.~\ref{R2vscap} plots the minimum rate requirement of the weak user against $C_{\textrm{max}}$ with $\sigma^2_{\mathbf{H}_1}=20$ dB, $\sigma^2_{\mathbf{H}_2}=1$ dB, $p_{\text{max}}=4$~W for various antenna array sizes. The range of values of $R_0$ is restricted by the assumption given in \eqref{eqnlimit}, however for larger MIMO arrays this restriction is reduced. We see that the asymptotic approach is optimal for any rate restraint whereas the suboptimal method deteriorates significantly when the rate requirement of the weak user increases and that the degree of the deterioration increases with $N$.

\begin{figure}[tbp]
\centering
\includegraphics[width=\columnwidth]{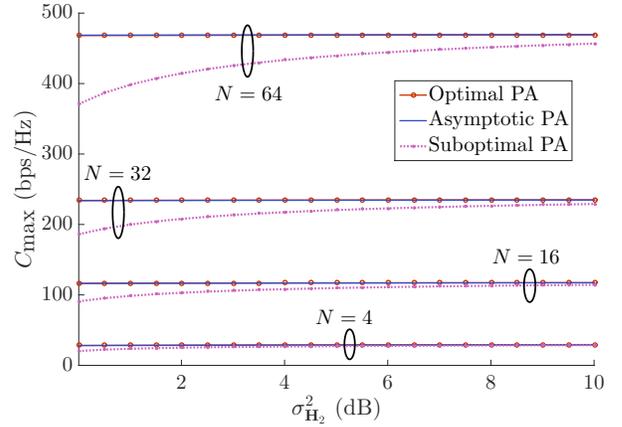}
\caption{Sum-capacity vs channel gain of weak user}
\label{sigmasqvscap}
\end{figure}

Fig. \ref{sigmasqvscap} plots the channel gain of the weak user against $C_{\textrm{max}}$, for $\sigma^2_{\mathbf{H}_1}=20$~dB, $p_{\text{max}}=4$~W, $R_0=2$~bps/Hz and various antenna array sizes. Again, the performance of the suboptimal method suffers for larger antenna arrays, most significantly in the case where the channel gain of the weak user is very small compared to that of the strong user, $\sigma^2_{\mathbf{H}_1}<< \sigma^2_{\mathbf{H}_2}$, which would happen when the strong user was very near to the base station while the weak user was very remote. As before, the asymptotic approach remains accurate in all cases.

Next we consider the computational complexity, which depends on the number of antennas (for which we will consider the case where $N_\mathrm{S}\neq N_i$), the number of iterations used to compute the expectations involved in the optimal and suboptimal methods $K$, and the number of bisections $M$ required for the optimal and asymptotic methods. 

The optimal bisection method is the most complex. It involves looping through the computation $M$ times and computing $C_1$ $K$ times in each loop to find the expectation. The complexity order of calculating $C_1$ is $\mathcal{O}(N_1!)$ since the most complex operation is taking the determinant of the $N_1\times N_1$ matrix $\left[\mathbb{I}_{N_1}+(\mathbb{I}_{N_1}+(p_2\mathbf{H}_1\mathbf{H}_1^\dag)^{-1})p_1\mathbf{H}_1\mathbf{H}_1^\dag\right]$ in \eqref{eqnc1} (recall that $\mathbf{H}_i\in\mathbb{C}^{N_i\times N_\mathrm{S}}$). The overall complexity order of this method is $\mathcal{O}(K M N_1 !)$, where we note that increasing $N_\mathrm{S}$ and $N_2$ does increase the complexity, but the complexity order is dominated by $N_1$. 

In comparison the asymptotic approach also loops over the capacity computation $M$ times but computes the capacity using the closed form in \eqref{eqnc1asy}, for which the complexity is invariant with respect to $N_\mathrm{S}$, $N_i$, $K$ and $M$, thus the overall complexity order of this method is $\mathcal{O}(M)$. 

Finally, the complexity of the suboptimal approach does not require looping through $M$ bisections, however it still involves computing the expectation over $K$ iterations of a computation involving the determinant of an $N_1 \times N_1$ matrix, thus it has complexity order $\mathcal{O}(K N_1!)$.

We note that the complexity order of the determinant computation can be reduced from $\mathcal{O}(N_1!)$ to as little as $\mathcal{O}(N_1^{2.81})$ using the methods in  \cite{complexitydet}[Theorem 6.6]. However, the implementation of these methods is beyond the scope of this paper. We have used the Matlab function \verb#det#, which relies on the $LU$ decomposition method for calculating the determinant and has complexity order $\mathcal{O}(N_1^3)$, which gives complexity orders $\mathcal{O}(K M N_1^3)$, $\mathcal{O}(M)$ and $\mathcal{O}(K N_1^3)$ for the respective methods.



\begin{figure}[tbp]
\centering
\includegraphics[width=\columnwidth]{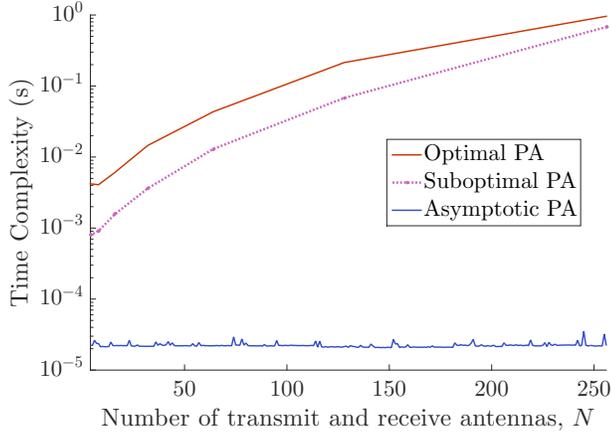}
\caption{Time complexity of power allocation algorithms}
\label{complex}
\end{figure}

We compare the time complexity of the three approaches for increasingly large antenna arrays in Fig.~\ref{complex}. Note that we fixed $K\!=\!10$ for the expectation calculations. Experimentation demonstrated that accurate results for the considered range of $N$ are observed if the number of bisections is at least \mbox{$M=13$} for $\epsilon=0.001$ ($\epsilon$ is used in Table~\ref{tab1}). With $K$ and $M$ fixed, the complexity of the optimal and suboptimal methods depends only on the number of antennas, as is corroborated by Fig.~\ref{complex}. In agreement with our calculations, the complexity of the asymptotic approach remains constant regardless of the size of the antenna array. 


\section{Conclusions}
We have used asymptotic analysis to extend the results of  \cite{sun2015ergodic} and demonstrated how best to allocate power resources to achieve optimal sum-capacity for an MM-NOMA system. We have demonstrated that the proposed asymptotic approach performs optimally for arbitrarily large antenna arrays while the accuracy of the suboptimal method of \cite{sun2015ergodic} decreases significantly with size for arrays larger than $4\times4$. Moreover, we have shown that the suboptimal method deteriorates in the cases of (i) low total power availability (ii) high minimum rate requirement at the weak user and (iii) significant difference between channel gains of users. The asymptotic method, on the other hand, agrees with the optimal method and is unaffected by these changes. Finally, we have demonstrated that the complexity of the asymptotic algorithm is lower than that of the optimal and suboptimal approaches regardless of array size. We conclude that the proposed power optimization method is superior for MM-NOMA.

\bibliographystyle{IEEEtran}
\bibliography{IEEEabrv,referencesnew}

\end{document}